\documentclass[aps,prd,11pt,superscriptaddress,notitlepage,longbibliography,nofootinbib,tightenlines,superscriptaddress]{revtex4-1}
\usepackage{natbib}
\usepackage{graphicx}
\usepackage{verbatim}
\usepackage{color}
\usepackage{bm}
\usepackage{amssymb,amsbsy,amsfonts,amsmath}
\usepackage{hyperref}
\hypersetup{colorlinks=true, linkcolor=darkred, citecolor=blue, linktoc=page}
\definecolor{darkred}{rgb}{0.8,0.1,0.1}

\newcommand{\eq}{\begin{equation}}
\newcommand{\eqe}{\end{equation}}

\def\tr{\text{tr}\,}
\def\trF{\text{tr}_{\rm F}\,}
\def\trA{\text{tr}_{\rm Adj}\,}
\def\Nc{N_{\rm c}}
\def\Nf{N_{\rm f}}
\def\mq{m_{q}}
\def\O{\mathcal{O}}
\def\LambdaQCD{\Lambda_{\rm QCD}}
\def\vec#1{{\bf #1}}

\hypersetup{
    colorlinks=true,       
    linkcolor=red,          
    citecolor=blue,        
    filecolor=magenta,      
    urlcolor=blue           
}

\newcommand{\bea}{\begin{eqnarray}}
\newcommand{\eea}{\end{eqnarray}}

\def\Th#1#2{\vartheta{\tiny\begin{bmatrix}{#1}\\{#2}\end{bmatrix}}}

\begin{document}

\preprint{INT-PUB-16-049}

\title{Exponential reduction of finite volume effects with twisted boundary conditions} 

\author{Aleksey Cherman}
\email{aleksey.cherman.physics@gmail.com}
\affiliation{Institute for Nuclear Theory, University of Washington, Seattle, WA 98195 USA}
\author{Srimoyee Sen}
\email{srimoyee08@gmail.com}
\affiliation{Department of Physics, University of Arizona, Tucson, AZ 85721 USA}
\author{Michael L. Wagman}
\email{mlwagman@uw.edu}
\affiliation{Institute for Nuclear Theory, University of Washington, Seattle, WA 98195 USA}
\affiliation{Department of Physics, University of Washington, Seattle, WA 98195 USA}
\author{Laurence G. Yaffe}
\email{yaffe@phys.washington.edu}
\affiliation{Department of Physics, University of Washington, Seattle, WA 98195 USA}

\begin{abstract}
    Flavor-twisted boundary conditions can be used for exponential reduction of
    finite volume artifacts in flavor-averaged observables
    in lattice QCD calculations with $SU(N_f)$ light quark flavor symmetry.
    Finite volume artifact reduction arises from destructive interference effects in a manner closely
    related to the phase averaging which leads to large $\Nc$ volume independence.
    With a particular choice of flavor-twisted boundary conditions,
    finite volume artifacts for flavor-singlet observables 
    in a hypercubic spacetime volume 
    are reduced to the size of finite volume artifacts 
    in a spacetime volume with periodic boundary conditions that is four times larger.
\end{abstract}
\maketitle

\section {Introduction} 
Many questions about QCD and related strongly coupled 4D quantum
field theories can only be systematically addressed using 
lattice gauge theory simulations.
Numerical calculations are necessarily performed in finite Euclidean
spacetime volumes, so extraction of physical quantities
of interest in a theory defined on $\mathbb R^4$ always involves
extrapolation to the infinite-volume limit as well as the continuum limit.
Performing this extrapolation with controlled errors requires
additional simulations using multiple lattices of varying size and/or
reliable independent knowledge of the volume dependence of observables.

This paper presents a technique to reduce
the size of finite volume artifacts for flavor singlet
observables in gauge theories with a
vector-like $SU(\Nf)_V$ flavor symmetry.%
\footnote
    {%
    Other proposals for exponential reduction of finite volume artifacts
    using related methods and other techniques have previously been explored in refs.~%
    \cite{Davoudi:2011md,Briceno:2013hya,Korber:2015rce}.
    }
We assume throughout that the lightest hadronic states in the theory
under study are flavor non-singlet mesons,
and also assume that the box size $L$ (and inverse temperature $\beta$)
are large compared to the inverse of the appropriate strong scale, $\LambdaQCD^{-1}$.
In particular,
our results are applicable to QCD in the isospin-symmetric limit
of degenerate up and down quark masses.
In a hypercubic box with sides of length $L$, our technique eliminates
the leading $O(e^{-m_{\pi}L})$ finite-volume effects,
where $m_{\pi}$ is the mass of the lightest meson.
The surviving residual finite volume artifacts in, for example,
hadron masses are exponentially smaller and
scale as $O(e^{-\sqrt 2 m_\pi L})$, as discussed below.
When applied to calculations of flavor-singlet observables
(i.e., observables invariant under the $SU(\Nf)_V$ symmetry)
in a spacetime volume $\mathcal{V}  = \beta L^3$,
with $\beta \ge L$,
our prescription leads to finite volume artifacts comparable to those
which arise with periodic boundary conditions in a 
larger spacetime box of volume
$
    \widetilde{\mathcal{V}}
    =
    4 \mathcal{V}\sqrt{(1{+}L^2/\beta^2)/2}
$.
Spectroscopy also benefits from reduced contamination 
from backwards-propagating thermal artifacts.
Finite volume effects associated with multi-hadron interactions
receive calculable modifications and are not generically exponentially reduced.

The technique we describe uses flavor-twisted boundary conditions (TBCs).
These have a long history in lattice QCD
\cite{Bedaque:2004kc,
Sachrajda:2004mi,
deDivitiis:2004kq,
deDivitiis:2004rf,
Guadagnoli:2005be,
Mehen:2005fw,
Tiburzi:2005hg,
Tiburzi:2006px,
Jiang:2006gna,
Simula:2007fa,
Hu:2007ts,
Jiang:2008te,
Jiang:2008ja,
Boyle:2008yd,
Aoki:2008gv,
Bedaque:2008hn,
Boyle:2012nb,
Brandt:2013mb,
Briceno:2013hya,
Bijnens:2014yya,
Korber:2015rce,
Lehner:2015bga,
Colangelo:2016wgs},
but
the main focus of most prior work has involved the use of TBCs
for valence quarks
to probe a finer set of momenta than are allowed by conventional
periodic boundary conditions (PBCs) for a given box size.
The utility of TBCs to reduce finite-volume artifacts in the context of lattice QCD was recently
explored in ref.~\cite{Briceno:2013hya}
(see also ref.~\cite{Korber:2015rce}).
Our results generalize some of the boundary conditions 
discussed in ref.~\cite{Briceno:2013hya},
and provide general symmetry-based arguments demonstrating that they reduce
finite-volume artifacts in generic flavor-singlet systems.
The basic technical ideas underpinning our proposal
are the freedom to view twisted boundary conditions as particular
choices for holonomies of background flavor gauge fields,
and the ability to write observables in a form where effects
of background gauge transformations are manifest.

To illustrate the essential concepts, let us consider QCD with a single
compactified direction, say 
the Euclidean time direction $x_4$.
Impose on the quark field $q(x)$, transforming in the
fundamental representation of the flavor symmetry group $SU(\Nf)_V$,
the periodicity condition
\begin{align}
    q(\vec{x},x_4+\beta)= \pm \Omega_4 \, q(\vec{x},x_4) \,.
\label{eq:singletwist}
\end{align}
Here $\beta$ is the circumference of the $x_4$ direction
and $\Omega_4$ is an element of $SU(\Nf)$ which is naturally viewed as an
$\Nf \times \Nf$ unitary matrix.
A global flavor transformation, $U \in SU(\Nf)_V$,
acts on quark fields as $q(x) \to U \, q(x)$ and
consequently has the effect of conjugating the periodicity condition,
$\Omega_4 \to U^\dag \Omega_4 U$.   
Hence, only the conjugacy class of $\Omega_4$ is physically significant.  
A generic choice of $\Omega_4$ breaks the $SU(N_f)_V$ flavor symmetry to its
Cartan subgroup $U(1)^{\Nf-1}$.  
As illustrated explicitly in Section~\ref{sec:massShift},
a generic holonomy $\Omega_4$ may induce finite-volume mass splittings
within the lightest meson multiplet, with non-trivial mixing
among the neutral mesons (those invariant under the Cartan subgroup),
but no mixing between mesons with different $U(1)^{\Nf-1}$ charges.

Three different perspectives on the ``twist'' $\Omega_4$ are useful.
One is to view $\Omega_4$ as defining a twisted boundary condition,
as introduced in Eq.~\eqref{eq:singletwist}.  
Alternatively, one may perform a field redefinition which makes the
(redefined) quark field strictly periodic (or anti-periodic) in $x_4$, 
at the cost of introducing a set of imaginary chemical potentials
$\mu = i\alpha /\beta$, with $e^{i\alpha}$ the eigenvalues of $\Omega_4$.
Finally, thanks to the well-known relation between imaginary
chemical potentials and background gauge fields,
one may also view the twist as the holonomy of a 
background $SU(\Nf)$ flavor gauge field $\mathcal{A}_{4}$,
$\Omega_4 = \mathcal{P} \, e^{i \oint dx_4 \> \mathcal{A}_4}$.

The conventional choice of Euclidean-time boundary conditions for
fermionic fields corresponds to selecting the minus sign in
Eq.~(\ref{eq:singletwist}) and setting $\Omega_{4} = 1_{\Nf}$.
With this choice,
the Euclidean path integral computes the thermal partition function,
$Z(\beta) \equiv \tr e^{-\beta H}$,  with $H$ the Hamiltonian.
Choosing $\Omega_4 \neq 1_{\Nf}$, and/or the plus sign in the periodicity
condition is also perfectly permissible,
but gives a functional integral which computes a \emph{twisted} partition
function.
For example, if one chooses the periodicity condition
$q(\vec{x},x_4+\beta)= + \Omega_4\, q(\vec{x},x_4) $ then
the Euclidean functional integral yields
\begin{align}
    \widetilde{Z}(\beta; \alpha) 
    \equiv
    \tr \bigl[ (-1)^F e^{-\beta H} e^{i \sum_{k=1}^{\Nf} \alpha_k \, Q_k} \bigr],
\label{eq:Ztilde}
\end{align}
where $F$ is total fermion number, 
$\{ Q_k \}$ are the conserved flavor charges which
count the net number of fermions of flavor $k$
(and represent the Cartan elements of the Lie algebra $\mathfrak{u}(\Nf)$),
and $\{ e^{i\alpha_k} \}$ are the eigenvalues of the twist $\Omega_4$.  

All eigenstates of the Hamiltonian contribute positively
in the thermal partition function $Z(\beta)$,
with their weights determined only by their energy.
In contrast,
in the twisted partition function $\widetilde{Z}(\beta,\alpha)$
states with different values of the commuting flavor charges $Q_k$
receive distinct phases.
There is value in considering twisted partition functions
even if one's primary interest involves thermal physics at finite $\beta$.
The symmetry-based cancellations inherent in $\widetilde{Z}$ allow one to
focus on the behavior of particular symmetry sectors of states by 
suitably dialing the boundary conditions.
Twisted partition functions become even more useful when
one's primary interest involves vacuum properties, and hence the
$\beta \to \infty$ limit.
As we will see, cancellations among states in the twisted partition
function can be arranged to eliminate the leading finite-$\beta$ dependence,
accelerating the convergence to the $\beta \to \infty$ limit.
This idea has been previously exploited in the literature on large
$\Nc$ twisted-Eguchi-Kawai reduction
\cite{GonzalezArroyo:1982hz,GonzalezArroyo:2010ss,Perez:2014sqa}
and in the literature on adiabatic circle compactifications
\cite{Unsal:2008ch,Unsal:2010qh,Poppitz:2013zqa,Cherman:2016hcd};
a number of our ideas were inspired by these works. 
As this paper was being finalized, 
ref.~\cite{Sulejmanpasic:2016llc} appeared with a systematic exploration
of twisted boundary conditions and volume dependence in $CP(N)$ and
$O(N)$ sigma models.

To be specific, consider the dependence on $\beta$ of the expectation
value of some flavor-singlet local observable, $\langle \O\rangle$, assuming that the spatial box size $L\gg\beta$.
Assume, for convenience, that
$m_{\pi} \ll \LambdaQCD$, so that the pions,
or more generally, the pseudo-Nambu-Goldstone bosons (pNGBs) of
$SU(\Nf)_L \times SU(\Nf)_R \to SU(\Nf)_V$
chiral symmetry breaking,
are much lighter than all other hadronic states,
and also assume, as stated earlier, that $\beta \gg \LambdaQCD^{-1}$.
In this regime,
chiral effective field theory provides a useful description of low
temperature dynamics.\footnote{
Chiral effective field theory has been used to study finite 
spacetime volume effects for a long time~\cite{Gasser:1986vb,Gasser:1987ah,Gasser:1987zq,Leutwyler:1987ak,Leutwyler:1992yt,Sharpe:1992ft,Colangelo:2003hf,Becirevic:2003wk,AliKhan:2003ack,Beane:2004tw,Beane:2004rf,Detmold:2004ap,Bedaque:2004dt,Arndt:2004bg,Colangelo:2005gd,Colangelo:2005cg,Bedaque:2006yi,Beane:2011pc}
.}
Leading finite-$\beta$ effects will come from pion loops,
with the chiral loop expansion controlled by the small
parameter $m_{\pi}^2/\left(4\pi f_{\pi}\right)^2 \ll 1$,
with $f_{\pi} \sim \LambdaQCD$.
The resulting chiral expansion for the expectation value
$\langle \mathcal O \rangle$ will have the form
\begin{align}
    \langle\O\rangle (\beta)
    =
    \O_0
    + \frac{m_{\pi}^2}{(4\pi f_{\pi})^2} \,
    \O_1 \big(m_{\pi} \beta, \Omega_4\big)
    + \cdots \,,
 \label{eq:finiteBetaO}
\end{align}
where $\O_0$ is a no-pion-loop contact term and
$\O_1$ is the one pion loop contribution.
This term will depend on the dimensionless ratio $m_\pi \beta$
as well as the the flavor twist $\Omega_4$,
and may be decomposed into a sum of terms coming from a given
winding number of the pion around the compact direction.
For large $\beta$,
winding number $n$ contributions must fall exponentially as
$O(e^{-|n| m_\pi \beta})$.
The key point is that background flavor gauge invariance constrains the
dependence of each such term on the holonomy $\Omega_4$.
If $\O$ is a flavor singlet,
then the contribution to $\O_1$ from winding number $n$ 
must involve an appropriate group invariant constructed from
$(\Omega_4)^n$.
Because the pNGBs transform in the adjoint representation of
$SU(\Nf)_V$, this invariant is simply the adjoint representation
trace.
Consequently, 
\begin{align}
    \O_1 \big(m_{\pi} \beta, \Omega_4\big)
    =
    \sum_{n=-\infty}^{\infty}
	f_{n}\big(m_{\pi} \beta\big) \, \trA [\Omega_4^n] \,,
 \label{eq:one-loop}
\end{align}
with $f_n(m_\pi\beta) \sim e^{-|n|m_\pi \beta}$ at large $\beta$
and the zero winding term $f_0$ being $\beta$-independent.
The representation (\ref{eq:one-loop}) shows that $\beta$-dependent 
one-pion-loop contributions necessarily involve a single adjoint trace
of the holonomy.
Note that adjoint representation traces ($\trA\!$) can always be
rewritten in terms of fundamental representation traces ($\trF\!$)
for the group $SU(\Nf)$,
namely $\trA \Omega = |\trF \Omega|^2 -1$.

Consequently, to eliminate the leading $\beta$ dependence one may choose
any flavor holonomy $\Omega_4$ for which $\trA \Omega_4 = 0$.
Such a choice of holonomy produces an exponential reduction in the
size of the finite-$\beta$ artifacts.
A particularly nice choice of the flavor holonomy will eliminate more
than just winding number $\pm 1$ contributions;
in the one-pion-loop result $\O_1$ it is possible to eliminate contributions
from all winding numbers of magnitude less than $\Nf{+}1$.
Specifically, let us define $\gamma \equiv e^{2\pi i /(\Nf+1)}$,
and choose the holonomy $\Omega_4$ to equal
\begin{align}
    \Gamma \equiv
    - \mathrm{diag}(\gamma, \gamma^2, \cdots, \gamma^{\Nf})\,.
\label{eq:vanishingadjoint}
\end{align}  
Up to an $SU(\Nf)$ similarity transformation,
this is a unique $SU(\Nf)$ matrix that obeys the relation
\begin{equation}
    \trA \Gamma^n =
    \begin{cases}
	0 \,, &  \mbox{for all $n \ne 0 \pmod {\Nf{+}1}$};
    \\
	\Nf^2{-}1 \,, & \mbox{for all $n$ divisible by $\Nf{+}1$}.
    \end{cases}
\label{eq:trGamma}
\end{equation}
We will refer to $\Gamma$ as the ``vanishing-adjoint-trace''
background flavor holonomy, and $\Omega_4 = \Gamma$ boundary
conditions as ``vanishing adjoint'' BCs.

Vanishing-adjoint boundary conditions with $\Nf=2$ have been previously
studied alongside other choices of twisted boundary conditions for
few-baryon systems in ref.~\cite{Briceno:2013hya}.
Vanishing-adjoint BCs were observed to remove the leading $O(e^{-m_\pi L})$
finite volume corrections to flavor-averaged baryon masses. However, the
vanishing-adjoint-trace mechanism and its implications for finite volume
artifact reduction in other systems were not detailed in ref.~\cite{Briceno:2013hya}.

When $m_\pi \ll \LambdaQCD$ and $L, \beta \gg \LambdaQCD^{-1}$,
the leading finite-temperature and finite spatial-volume effects are
accurately described by one-loop chiral effective theory (EFT).
If in addition $L \gg \beta$, then
the argument above shows that the vanishing-adjoint
twisted boundary condition \eqref{eq:vanishingadjoint}
eliminates the first $\Nf$ finite-$\beta$ corrections and
leads to a dramatic decrease of the magnitude of finite-$\beta$
artifacts in the one-loop contribution to $\langle \O\rangle$,
from $O(e^{-\beta m_{\pi}})$ to $O(e^{-(\Nf+1)\beta m_{\pi}})$.
In section \ref{sec:loops} we discuss the extent
to which this feature persists at higher loops.

If multiple compactified directions are of comparable size,
then the analysis is somewhat more involved.
To clarify the effects of choosing vanishing-adjoint boundary
conditions on the size of both finite-volume and finite-temperature
artifacts, 
in the remainder of this paper we explicitly compute the volume and
temperature dependence of several flavor-singlet physical quantities
in QCD-like theories compactified on $T^4$ with twisted boundary conditions
in all directions.

Throughout this paper we assume that:
(\emph{i})
    the chosen numbers of colors ($\Nc$) and (complex) fundamental representation
    flavors ($\Nf$) place the theory (on $\mathbb R^4$) in a confining phase;
(\emph{ii})
    there is an unbroken $SU(\Nf)_V$ flavor symmetry acting on the
    lightest quarks; and
(\emph{iii})
    the mass $\mq$ of these lightest quarks is sufficiently small
    so that the lightest hadrons are mesons, not glueballs
    (i.e., $\mq$ is not large compared to $\LambdaQCD$).
Within the following three sections examining the free energy,
pion propagator, and pion mass shift we will make one further
simplifying assumption,
namely that the theory is close to the chiral limit, $\mq \ll \LambdaQCD$.
This will enable us to perform quantitative calculations using chiral
perturbation theory.
But, as we discuss in the concluding section,
the utility of flavor-twisted boundary conditions for reducing
finite volume effects does not require that the theory be parametrically close
to the chiral limit.
Rather, what is required is that the $SU(\Nf)$ multiplet of quarks
not be so heavy that the lightest hadrons become flavor singlet states
instead of flavor adjoints.

\section {Free energy}  

As a first example, we consider the free energy for a gauge
theory with $\Nf$ light degenerate fundamental representation
quarks on a large four-torus.
In addition to its intrinsic interest, evaluation of the
finite volume free energy allows one to compute finite volume
corrections to the chiral condensate by taking a derivative with
respect to the quark mass $\mq$.
We denote the circumferences of the fundamental cycles of the $T^4$
by $L_\mu$, $\mu=1,{\cdots},4$.
One may regard the torus as having a spatial volume 
$V = L_1 L_2 L_3$ and temporal extent $\beta = L_4$.
We impose twisted boundary conditions (TBCs) on the quark fields,
\begin{equation}
  q(x_\mu + L_\mu) = \Omega_\mu \, q(x_\mu) \,,
  \label{quarkTBCdef}
\end{equation}
and require that the flavor holonomies $\Omega_\mu$ be mutually commuting
$SU(\Nf)$ matrices.
For generic choices of commuting flavor holonomies,
these boundary conditions explicitly break
the $SU(\Nf)_L \times SU(\Nf)_R$ chiral symmetry (of the massless theory)
down to the maximal Abelian subgroup $U(1)^{\Nf-1}_L \times U(1)^{\Nf-1}_R$,
with the size of the symmetry breaking scaling as $1/L_{\mu}$.
Choosing a diagonal basis, without loss of generality,
we define the twist angles  $\{ \alpha_{\mu}^{a} \}$ via
\begin{equation}
    (\Omega_{\mu})^{ab} = e^{i \alpha^a_{\mu}} \, \delta^{ab}.
\label{quarkholonomydef}
\end{equation}
Here and henceforth
we use lower case letters $a,b,{\cdots} = 1,{\cdots},\Nf$
to denote fundamental representation $SU(\Nf)_V$ indices, 
and upper case letters $A, B, {\cdots} = 1,{\cdots},\Nf^2{-}1$
to denote adjoint representation indices.

When $\beta \LambdaQCD^{-1}$ and $L \LambdaQCD^{-1}$ are both large,
the twisted free energy will be dominated by the lightest modes in the system,
the pNGB `pions'.
A given pion $\pi^{ab}$ receives a twist angle
$\alpha^{ab}_{\mu} \equiv \alpha^{a}_{\mu} - \alpha^{b}_{\mu}$.
Equivalently, using an adjoint representation basis the pion boundary
conditions read
\begin{equation}
  \pi^A(x_\mu{+}L_\mu) = \sum_{B} (\Omega_{\mu})_{\;\;B}^{A}\,\pi^B(x_\mu) \,.
\end{equation}
Even though the adjoint representation is a real representation,
it will be convenient to use a complex basis
which diagonalizes the $U(1)_V^{\Nf-1}$ Cartan subgroup.
The complex conjugate of the basis element with index $A$
will be a basis element (which may be the same or different)
whose index we will denote as $\bar A$.%
\footnote
    {%
    Under the adjoint action of the Cartan subgroup,
    there are $\Nf{-}1$ basis elements, corresponding to neutral
    Nambu-Goldstone bosons, which are invariant and have $\bar A = A$,
    while $\Nf(\Nf{-}1)$ basis elements, corresponding to
    charged Nambu-Goldstone bosons, are non-invariant and have
    $\bar A \ne A$.
    }
Given such a choice of basis,
the adjoint representation twists are diagonal,
\begin{equation}
  (\Omega_{\mu})_{\;\;B}^{A} = e^{i \alpha_\mu^A} \, \delta^{A}_{\;\;B} \,,
\end{equation}
with
$\alpha_\mu^A \equiv 2\; \tr[t^A \, \alpha_\mu]$ and
$\alpha_{\mu} \equiv \| \alpha^{ab}_{\mu}\|$.
The adjoint basis matrices $\{ t^A \}$ are chosen to satisfy
\begin{equation}
  \tr \big(t^A t^{\bar B} \big) = \tfrac{1}{2} \, \delta^A_{\;\;B}\,.
\end{equation}
Having chosen a complex basis,
it is helpful to use upper and lower indices to distinguish complex conjugation
of generators, and to also define
\begin{equation}
    g^{AB} \equiv 2\, \tr \big(t^A t^B \big) = \delta^{A \bar B} \,,
    \hspace{20pt}
    g_{AB} \equiv 2\, \tr \big(t^{\bar A}t^{\bar B} \big) = \delta^{\bar A B} \,.
\end{equation}
These function as the components of our metric (and its inverse)
in the $SU(\Nf)_V$ Lie algebra.
We will need symmetric structure constants in this basis,
redundantly defined as
\begin{equation}
  d^A_{\;\;BC}
  =
  2\, \tr\big( t^A \big\lbrace t^{\bar B}, t^{\bar C}\big\rbrace \big),
  \hspace{20pt}
  d_A^{\;\;BC}
  =
  2\, \tr\big(  t^{\bar A} \big\lbrace t^{B}, t^{C}\big\rbrace \big) \,.
\end{equation}
(Antisymmetric structure constants are defined analogously but will not be needed below.)

Imposition of twisted boundary conditions is equivalent to
working with periodic fields in the presence of background $SU(\Nf)_V$
gauge fields given by
$(\mathcal{A}_\mu)^{ab} =  \alpha^a_{\mu}\, \delta^{ab}/{L_\mu}$.
We therefore define a background flavor covariant
derivative  $D_\mu = \partial_\mu + i \mathcal{A}_{\mu}$.
The Euclidean chiral Lagrangian which describes the low-energy dynamics
then takes the form 
\begin{align}
   \mathcal{L}
   &=
   f_\pi^2 \> \tr\!\big[
       D_\mu \Sigma \, D_\mu \Sigma^\dagger
       - 2B (\mq^\dagger \, \Sigma + \Sigma^\dagger \mq)
   \big] + \cdots \,,
\label{eq:LpiTBC}
\end{align}
up to four-derivative and higher terms. 
(For comparison to the literature, note that our conventions correspond to
$f_\pi \approx 46 \text{ MeV}$ for $N_f=2$ QCD.)
Writing
$
    \Sigma
    = \exp({i \pi}/{f_\pi})
    \equiv \exp({\sum_A i \pi^A t^{\bar A} }/{f_\pi})
$
and expanding in powers of the pion field $\pi$ gives
\begin{align}
   \mathcal{L}
   &=
   \tr (D_\mu\pi D_\mu\pi)
   + m_\pi^2 \, \tr(\pi^2)
   + \tfrac{1}{6} f_\pi^{-2} \,
       \tr( \pi D_\mu \pi \pi D_\mu \pi - \pi^2 D_\mu \pi D_\mu \pi )
   - \tfrac{1}{12} \, m_\pi^2f_\pi^{-2} \, \tr(\pi^4)
    + \cdots
   \,,
\end{align}
where we used $m_{\pi}^2 \equiv 2B\,\mq$
and have suppressed terms of order $\pi^6$ and higher.
 
The logarithm of the twisted partition function defines
the twisted free energy density,
$
    \widetilde{\mathcal{F}}(\beta,V; \Omega_\mu)
    \equiv
    - (\ln \widetilde{Z})/(\beta V)
$.
For light pions and large volumes (compared to the scale $f_\pi$),
the free energy is dominated by freely-propagating pions.
To lowest order
$
   \widetilde{Z} =
   \left[ \det \!\big( -D_\mu D^\mu+ m_\pi^2 \big) \right]^{-1/2}
$
and hence 
%
\begin{align}
    \widetilde {\mathcal{F}}(\beta,V;\Omega)
    &=
    \frac{1}{2\beta V}\sum_A \sum_{n_\mu\in\mathbb{Z}^4}
    \ln\left[ (\alpha_\mu^A + 2\pi n_\mu)^2 L_\mu^{-2} + m_\pi^2 \right] .
\label{eq:fdef}
\end{align}
This expression is UV divergent and requires regularization
and renormalization.
To extract the physical $\beta$ and $L$ dependent result it is convenient
to write
$
    \ln x = \lim_{s\to0} dx^s/ds
$
and subtract the corresponding decompactified $\beta = L_i = \infty$ expression.
Hence,
\begin{align}
    \widetilde {\mathcal{F}}(\beta,V;\Omega)
    &=
    \tfrac{1}{2}\sum_A
	\tfrac{\partial}{\partial s} \, T(s,\alpha_{\mu}^A) \Big|_{s=0}\,,
\end{align}
where $T(s,\alpha_{\mu})$ is the regularized tadpole sum,
\begin{align}
    T(s,\alpha_\mu) &\equiv
    \frac 1{\beta V}
    \sum_{n_\mu\in\mathbb{Z}^4}
	\left[ ( \alpha_\mu + 2\pi n_\mu)^2 L_\mu^{-2} + m_\pi^2 \right]^{-s}
    - \int\frac{d^4k}{(2\pi)^4} \> {(k^2+m_\pi^2)^{-s}} \,.
\end{align}
Algebraic manipulations (similar to those in, e.g., ref.~\cite{Sachrajda:2004mi})
allow one to put $T(s,\alpha_{\mu})$ into a more useful form,
\begin{align}
    T(s,\alpha_\mu) 
    &=
    {\Gamma(s)}^{-1}
    \int_0^\infty dz\> z^{s-1} \, e^{-z m_\pi^2}
	\bigg[
	    \frac{1}{\beta V}
	    \sum_{n_\mu\in\mathbb{Z}^4}
	    e^{ - z ( \alpha_\mu + 2\pi n_\mu)^2 L_\mu^{-2} }
	    - \int \frac{d^4k}{(2\pi)^4} \> e^{-z k^2}
	\bigg]
\nonumber
\\ &=
    {\Gamma(s)}^{-1}
    \int_0^\infty dz\> z^{s-1} \, e^{-z m_\pi^2}
    \bigg[
    \prod_\mu
	\frac{1}{L_\mu}
	\Th{\alpha_\mu/2\pi}{0}\big( {4\pi i z}{L_\mu^{-2}} \big)
	- \int_0^\infty \frac{dk\>k^3}{8\pi^2}\;e^{-zk^2}
    \bigg]
\nonumber
\\ &=
    {\Gamma(s)}^{-1}
    \int_0^\infty dz\> z^{s-1} \, e^{-z m_\pi^2}
    \bigg[
    \prod_\mu
    \frac{1}{\sqrt{4\pi z}}\;
    \Th{0}{\alpha_\mu/2\pi}\big( {-}{L_\mu^2}/(4\pi i z) \big)
    - \frac{1}{(4\pi z)^2}
    \bigg]
\nonumber
\\
    &=
    \frac{m_\pi^2}{2\pi^2\Gamma(s)} \, {(2 m_\pi)^{-s}}
    \sum_{n_\mu \in \mathbb{Z}^4}\!\!\!{\vphantom{\Big|}}^\prime \>
    e^{i n_\mu \alpha_\mu}\;
    |n L|^{s-2} \, K_{2-s}\left(m_\pi |n L|\right),
\label{eq:tadpolesum}
\end{align}
where
$|nL| \equiv \big(\sum_\mu n_\mu^2 L_\mu^2\big)^{1/2}$.
In the final expression \eqref{eq:tadpolesum},
the prime on the sum is an indication to omit the
$n_{\mu} = (0,0,0,0)$ term, and
$K_\nu(z)$ is the usual modified Bessel function.
In the second line we recognized the appearance of a
$\vartheta$-function with characteristics,
$
    \Th{\alpha}{\beta}(\tau)
    \equiv
    \sum_{n \in \mathbb{Z}} e^{2\pi i n\beta} e^{\pi i \tau (\alpha+n)^2}
$,
which was converted to the form in the third line using the modular
$S$ transformation
$
    \Th{a}{b}(\tau)
    =
    \sqrt{{i}/{\tau}}\>e^{-2\pi i ab} \,
    \Th{-b}{a}\big( -{1}/{\tau} \big)
$.
With the result \eqref{eq:tadpolesum} in hand,
computing the free energy is straightforward.
One finds
\begin{subequations}\label{eq:twistedFresult}%
\begin{align}
  \widetilde{\mathcal{F}}(\beta,V;\Omega)
  &=
    \sum_{n_\mu \in \mathbb{Z}^4}\!\!\!{\vphantom{\Big|}}^\prime \>
    m_\pi^2 \, f^{(1)}_{1}(|n L|) \>
    \trA[\Omega^{n}] \,,
\label{eq:twistedFResult}
\end{align}
where
$\Omega^{n} \equiv \Omega_1^{n_1}\Omega_2^{n_2}\Omega_3^{n_3}\Omega_4^{n_4}$
and 
\begin{align}
    f^{(1)}_{1}(|nL|)
    \equiv
    K_2\left( m_\pi |nL| \right) \big/ (2\pi \, |nL|)^{2} \,.
\end{align}
\end{subequations}
As advertised,
the flavor-singlet twisted free energy $\widetilde{\mathcal{F}}$ depends on
the holonomies $\Omega_\mu$ via an adjoint representation trace,
and at one-loop order in the chiral expansion only single-trace terms occur.
(We discuss higher-loop contributions in Section~\ref{sec:loops}.)

We now examine implications of the result \eqref{eq:twistedFresult}.
First, as a check, note that expression \eqref{eq:twistedFresult},
when evaluated with $\Omega_{\mu} = {\mathbf 1}$,
reduces in the chiral limit
to the correct infinite-volume massless Bose gas result,
\begin{align}
    \widetilde{\mathcal{F}}(\beta,\infty;\Omega_{\mu} = \mathbf{1})\big|_{m_\pi=0}
    =
    (\Nf^2 {-} 1) \, \frac{\pi^2}{90 \beta^4} \,.
\end{align}
Alternatively, if one takes $L_i \gg \beta$
so the spatial boundary conditions can be ignored,
and sets $\Omega_4 = \Gamma$, then one finds
\begin{align}
    \widetilde{\mathcal{F}}(\beta,\infty; \Omega_4 = \Gamma)
    &=
    \frac{m_\pi^2 \, (\Nf^2{-}1)}{2\pi^2 (\Nf{+}1)^2 \beta^2} \,
    \sum_{n\ge 1} n^{-2} \, K_2\big(n (\Nf{+}1) \beta \, m_{\pi}\big) \,.
\label{eq:twistedfreeenergy}
\end{align}
Using the Bessel function asymptotics,
$K_\nu(z) \sim \sqrt{\frac{\pi}{2z}}e^{-z}$,
one sees that the vanishing of adjoint traces of the flavor holonomy
(in this single compactified dimension regime)
has led to the advertised elimination of all finite $\beta$ contributions
involving winding numbers which are non-zero modulo $\Nf{+}1$.
In the chiral limit,
expression \eqref{eq:twistedfreeenergy} reduces to
\begin{align}
    \widetilde{\mathcal{F}}(\beta,\infty; \Omega_4 = \Gamma ) \big|_{m_\pi=0}
    =
    (\Nf^2 {-} 1) \, \frac{\pi^2}{90 (\Nf{+}1)^4\beta^4} \,.
\end{align}
We comment on the interpretation of this limiting result below.
Returning to the general one-loop result \eqref{eq:twistedFresult},
the following observations may be made:
\begin{itemize}
\item
    Imposing vanishing-adjoint BCs in all directions,
    $\Omega_{\mu} = \Gamma$, eliminates all leading $O(e^{-m_\pi L_\mu})$
    finite size free energy corrections to $\widetilde{F}$,
    and consequently also to thermodynamic derivatives
    such as the flavor-averaged chiral condensate.
\item
    Imposing vanishing-adjoint BCs in all directions does not
    remove all next-to-leading finite size corrections
    (arising from $n \cdot n = 2$ terms),
    such as the winding number $n_\mu = (1,-1,0,0)$ contribution.
    Nevertheless, vanishing-adjoint BCs reduce
    finite volume free energy corrections by exponentially large factors.
    In a hypercubic box, for example, the leading FV correction changes from
    $O(e^{-m_\pi L})$ with the usual periodic boundary conditions to
    $O(e^{-\sqrt{2}m_\pi L})$ with our vanishing-adjoint boundary conditions.
\item
    When finite-size artifacts from a single compactified dimension
    dominate (e.g., when $\beta \ll L_i$),
    then using the vanishing adjoint holonomy $\Gamma$ for this dimension
    removes the first $\Nf$ finite volume corrections from the
    one-loop free energy leaving
    $O(e^{-(\Nf+1)m_\pi \beta})$ finite-$\beta$ artifacts.
    The one-loop free energy in this regime is given by expression
    \eqref{eq:twistedfreeenergy}, which is precisely the
    free energy for free massive pions on an enlarged circle
    of size $(\Nf{+}1)\beta$.
\item
    For large $\Nf$
    (assuming one suitably scales $\Nc$ to retain asymptotic freedom
    and chiral symmetry breaking),
    pions are wholly insensitive to the thermal compactification,
    up to corrections which vanish as inverse powers of $\Nf$ in
    the chiral limit and fall exponentially with $\Nf$ for non-zero mass.
    The mechanism behind this `pionic volume independence'
    is essentially the same as in the more familiar case of
    large $\Nc$ volume independence.

\end{itemize}

Further discussion of extensions of the above features,
including higher loop contributions with flavor twisted boundary conditions
and connections between  flavor-twisted boundary conditions with large $\Nf$
and large $\Nc$ volume independence,
is postponed until section \ref{sec:loops}.

\section {Pion propagator}  

We now examine the effect of flavor-twisted boundary conditions
on the pion propagator.
Hadron propagators used for lattice QCD spectroscopy are 
two-point correlation functions, typically Fourier transformed in ``space'',
and hence depending on a spatial three-momentum and Euclidean time.
Information on the particle spectrum is extracted from their
exponential fall-off for large Euclidean time separations.
If $t$ is the time separation of the two insertions,
then the propagator of, for example, pions contains both
``forward'' contributions proportional to $e^{-m_\pi t}$
as well as ``backwards'' contributions
proportional to $e^{-m_\pi(\beta -t)}$.
These backwards-propagating thermal artifacts
effectively limit the useful time separations when
fitting correlation functions to half the Euclidean time extent.

Focusing, once again, on sufficiently large spacetime volumes and
small pion masses, the pion propagator can be evaluated using
chiral EFT.
The effect of a background flavor holonomy,
or flavor twisted boundary conditions,
is to shift the allowed values of momenta.
At tree level,
the (spacetime Fourier-transformed) pion propagator is
\begin{align}
  C(n_{\mu}; \Omega)^A{}_{B}
    \equiv \mathcal {F.T.} \, \langle \pi^A(0) \pi^{\bar B}(x) \rangle
    = \frac{\delta^A{}_{B}}{ N^A \cdot N^A  + m_\pi^2} \,,
\label{Cpidef}
\end{align}
where $n_{\mu} \in \mathbb{Z}^4$ and
$N^{A}_{\mu} \equiv (2\pi n_{\mu} + \alpha^{A}_{\mu})/L_{\mu}$ are
the discrete allowed wavevectors.
One may evaluate the discrete Fourier transform needed
to compute the position space pion propagator
$C^{A}{}_{B}(x_{\mu}; \Omega)$
using the same manipulations described above
for the tadpole integral \eqref{eq:tadpolesum}.
One finds,
\begin{align}
  C(x_{\mu}; \Omega)^A{}_{B}
    &=
    \frac{1}{\mathcal{V}} \sum_{n_{\mu} \in \mathbb{Z}^4} \>
    \frac{e^{-i x \cdot N^{A}}}{ N^A \cdot N^A + m_\pi^2}\; \delta^A{}_{B}
\nonumber
\\ &
    =
    \frac{1}{\mathcal{V}} \sum_{n_{\mu} \in \mathbb{Z}^4}\>
    \int_{0}^{\infty} dz \> e^{-z (N^A\cdot N^A  + m_\pi^2)}  \,
    e^{-i x \cdot N^{A}}\; \delta^A{}_{B}
\nonumber
\\ &
    =
    \frac{1}{\mathcal{V}}
    \int_{0}^{\infty} \!\! dz \> e^{-z m_{\pi}^2}
    \prod_{\mu=1}^{4}
	\Th{\alpha^{A}_{\mu}/2\pi}{-x_{\mu}/L_{\mu}}
	\big({4\pi i z}{L_{\mu}^{-2}}\big)\; \delta^A{}_{B}
\nonumber
\\ &
    =
    \int_{0}^{\infty} \!\! dz \>  
    \frac{e^{-z m_{\pi}^2}}{(4\pi z)^2}
    \prod_{\mu=1}^{4}
	\Th{x_{\mu}/L_{\mu}}{\alpha^{A}_{\mu}/2\pi}
	\big(-{L_{\mu}^2}/(4\pi i z)\big)\; \delta^A{}_{B}
\nonumber
\\ &=
    \frac{m_{\pi}}{4\pi^2}
    \sum_{n_{\mu} \in \mathbb{Z}^4}
    e^{i \alpha_{\mu}^{A} n_{\mu}} \,
    \frac{K_1\big(m_{\pi} |X(n)|\big)}{|X(n)|} \; \delta^A{}_{B} \,,
\label{eq:CA}
\end{align}
where $X_{\mu}(n) \equiv x_{\mu} + n_{\mu} L_{\mu}$
and $|X(n)| \equiv \sqrt{X(n) \cdot X(n)}$. 
Note there is no summation of repeated Lorentz indices implied in $n_\mu L_\mu$.
The result \eqref{eq:CA} is precisely a sum-of-images
representation of the periodic pion propagator, and could have
been written directly without starting from the Fourier representation.
The effect of the background holonomy is merely to insert appropriate
phase factors, depending on the winding numbers $n_\mu$, in each term.
The propagator \eqref{eq:CA} depends, of course, on the specific
pion flavor $A$ and its corresponding twist angles $\alpha^A_\mu$.
To obtain a flavor-singlet quantity,
we define the flavor-averaged pion propagator
\begin{align}
    C_\pi(x_{\mu}; \Omega)
    &\equiv
    \frac{1}{\Nf^2{-}1}\sum_A C(x_{\mu};\Omega)^{A}_{\;\;A}
    =
    \frac{m_{\pi}}{4\pi^2(\Nf^2{-}1)}
    \sum_{n_{\mu} \in \mathbb{Z}^4}
    \frac{K_1\left(m_{\pi} |X(n) |\right)}{|X(n)|}\, \trA [\Omega^{n}]  \,.
\label{eq:averageCpisum}
\end{align} 
When $m_\pi |x| \gg 1$ (and $m_{\pi} L_\mu \gg 1$),
the Bessel functions can be approximated by their asymptotic forms and
one obtains
\begin{align}
    C_\pi(x_{\mu}; \Omega)
    &\sim
    \frac{m_\pi^2 \, e^{- m_\pi |x|}}{(2\pi m_\pi |x|)^{3/2}} 
    +
    \frac{m_\pi^2}{\Nf^2 {-}1}
    \sum_{n_\mu\in \mathbb{Z}^4}\!\!\!{\vphantom{\Big|}}^\prime \>
    \frac{e^{-m_\pi|X(n)|}}{(2\pi m_\pi |X(n)|)^{3/2}}\,
    \trA[\Omega^n] \,.
\end{align} 
The second term is the finite volume (and finite $\beta$) correction.
Backwards-propagating thermal artifacts are most easily seen in
terms in this sum with $n_4 \neq 0$ and $n_{i} = 0$.
Supposing $x_4 \gg \mathrm{max}(|x_i|)$, so that
$|X(n)| \approx |x_4 + n_4 \beta|$, these terms become
\begin{align}
\label{eq:thermalnoise}
    C_\pi(x_{\mu}; \Omega)
    &\sim
    \frac{m_\pi^2 \, e^{- m_\pi |x_4|}}{(2\pi m_\pi |x_4|)^{3/2}}
    + 
    \frac{m_\pi^2}{\Nf^2{-}1} \sum_{n_4 \ne 0}
    \frac{e^{- m_\pi |x_4-n_4\beta|}}{(2\pi m_\pi |x_4 {-} n_4 \beta|)^{3/2}} \,
    \trA[\Omega^{-n_4} ]
    +\cdots \,,
\end{align}
and the $n_4 = 1$ term proportional to $ e^{-m_{\pi}|x_4-n_4 \beta|}$
provides the leading backwards-propagating thermal artifact.
If one imposes periodic boundary conditions on the pions, $\Omega_{\mu} = 1$,
then forward and backward propagating contributions are comparable at $x_4 \sim \beta/2$.
If one instead chooses $\Omega_4 = \Gamma$,
and neglects the spatial holonomies
(either because they are set to unity, or because $L \gg \beta$),
then the result \eqref{eq:thermalnoise},
combined with the vanishing adjoint traces \eqref{eq:trGamma},
shows that all backwards-propagating contributions
(as well as fully wrapped forward contributions)
are eliminated except those
which involve wrappings by integer multiples of $\Nf{+}1$.

In the chiral limit, the flavor-averaged pion propagator
\eqref{eq:averageCpisum} reduces to
\begin{equation}
    C_\pi(x_{\mu}; \Omega) \big|_{m_\pi = 0}
    =
    \frac 1{\Nf^2{-}1}
    \sum_{n_{\mu} \in \mathbb{Z}^4}
    \frac{\trA [\Omega^{n}]}{4\pi^2 \, X(n)^2}\, \,.
\end{equation}
If one chooses $\Omega_4 = \Gamma$, then
in the regime $L \gg \beta \gg |x|$ this reduces to
\begin{equation}
    C_\pi(x_{\mu}; \Omega_4=\Gamma) \big|_{m_\pi = 0}
    =
    \frac 1{4\pi^2 x^2}
    +
    \frac 1{2\pi^2}
    \sum_{n\ge 1}
    \frac 1{[(\Nf{+}1)\, n \beta]^2}
    =
    \frac 1{4\pi^2 x^2}
    +
    \frac 1{12\, (\Nf{+}1)^2 \beta^2} \,,
\label{eq:Cpichiral}
\end{equation}
showing that the residual finite-$\beta$ correction is suppressed by
a factor of $1/(\Nf{+}1)^2$.

Alternatively, if $L \lesssim \beta$ and one chooses $\Omega_i = \Gamma$
(in hopes of eliminating both finite $\beta$ and finite spatial volume effects),
then the result is more involved.
The largest $\Nf$ backwards-propagating thermal artifacts
(those coming from terms in the sum with $|n_4| \le \Nf$ but $n_{i} = 0$)
are eliminated, but some backwards-propagating terms with $n_{i} \neq 0$,
such as $n_{\mu} = (0,0,1,-1)$, survive.
This precisely parallels the above-discussed situation with the free energy.
When $\Omega_{\mu} = \Gamma$,
finite-spacetime-volume effects in flavor averaged
two-point functions in a hypercubic spacetime box of volume $\mathcal{V}$ 
resemble those of a theory with untwisted boundary conditions
in a box of volume $4\mathcal{V}$.

\section {Pion mass shift}
\label{sec:massShift}

The tree-level pion propagator receives corrections arising from interactions.
For sufficiently light pseudo-Nambu-Goldstone bosons, these corrections
may be calculated using chiral EFT in a finite volume
with twisted boundary conditions
\cite{Sachrajda:2004mi,Jiang:2006gna,Bijnens:2014yya,Colangelo:2016wgs}.
The explicit form of finite volume corrections to the masses
of individual pNGBs can be rather involved in the presence of TBCs.
However, finite volume corrections to the average pNGB mass --- a flavor-singlet quantity ---
can only depend on the holonomy via traces,
analogously to Eq.~\eqref{eq:one-loop}, and will therefore be exponentially
reduced from $O(e^{-m_\pi L})$ to $O(e^{-\sqrt{2}m_\pi L})$
(for a hypercubic box)
by vanishing-adjoint twisted boundary conditions.

This general argument may be verified by explicit calculation.
With interactions included, and twisted boundary conditions,
the pion propagator can become non-diagonal in flavor,
\begin{equation}
  C(k_\mu;\Omega)^A_{\;\;B}
    \equiv
    \mathcal {F.T.} \,\langle \pi_A(0) \pi_{\bar B}(x) \rangle
    =
    \bigl[\left(
	\mathbf P^2 + m_\pi^2 \, \mathbf 1 + \mathbf \Sigma
      \right)^{-1}\bigr]^A_{\;\;B} \,,
\end{equation}
where $k_\mu\in\mathbb{Z}^4$,
$\mathbf 1 \equiv \| \delta^{A}_{\;\;B} \|$ and
$\mathbf P_\mu  \equiv \| P^A_\mu \> \delta^{A}_{\;\;B} \|$,
with $P_\mu^A = (2\pi k_\mu + \alpha_\mu^A)/L_\mu$
the incoming pNGB momentum.
The self-energy
$\mathbf\Sigma \equiv \| \Sigma(k)^{A}_{\;\; B}\|$
receives one-loop contributions from a single tadpole-type diagram,
but with momentum-dependent vertex factors.
The unbroken $U(1)^{\Nf-1}_V$ symmetry implies that there is no
mixing of charged pions;
if $A \ne \bar A$ then $C^{A}_{\;\;B} \propto \delta^A_{\;\;B}$.
But with generic commuting flavor holonomies no symmetry 
prevents the $\Nf{-}1$ uncharged pNGBs from mixing.
The finite-volume one-loop self-energy, with twisted boundary conditions, reads
\begin{align}
  \Sigma(k)^{A}_{\;\;\;B}
    &=
    -\frac{1}{24f_\pi^2\beta V}
    \sum_{n_\mu\in\mathbb{Z}^4}\sum_{C,D}
    \frac{\delta^C_{\;\;D}}{N^C \cdot N^C + m_\pi^2} \, \times
\nonumber \\
    &\qquad \>\>
    \left\lbrace
        \big(
        \tfrac{2}{\Nf}\, \delta^A_{\;\;B} \, \delta^D_{\;\;C}
	    + d^A_{\;\;BE} \, d_C^{\;\;DE} \big)
	\left( m_\pi^2 +  2P^A \cdot P^A + 2N^C \cdot N^C\right)
    \right.
\nonumber \\
    &\qquad +
	\big(
        \tfrac{2}{\Nf}\, \delta^{A}_{\;\;C}\, \delta^D_{\;\;B}
        + d^A_{\;\;CE}\, d_{B}^{\;\;D E} \big)
	\left( m_\pi^2 - P^A \cdot P^A - N^C \cdot N^C - 6 P^A \cdot N^C \right)
\nonumber \\
    &\qquad +
    \left.
    \big(
	\tfrac{2}{\Nf}\, g^{A D}\, g_{B C}
        + d^E_{\;\;BC}\, d_E^{\;\;AD} \big)
    \left( m_\pi^2 - P^A \cdot P^A - N^C \cdot N^C + 6P^A \cdot N^C \right)
    \right\rbrace,
\label{1loopFVdiag}
\end{align}
with $N_\mu^C \equiv (2\pi n_\mu + \alpha_\mu^C)/L_\mu$
the internal loop momentum.
Terms involving numerators such as
$P^A \cdot N^C$,
linear in the pion loop momentum,
vanish by spacetime symmetries in infinite volume,
or in cubic compactifications with ordinary periodic boundary conditions.
However,
such terms need not vanish with generic flavor twisted boundary conditions,
and lead to a linear momentum dependent shift in the self-energy
\cite{Jiang:2006gna,Bijnens:2014yya,Colangelo:2016wgs}.
As noted in these references,
properly extracting the pion mass from the propagator requires care
in the presence of TBCs.
Near the poles of the propagator
(when analytically continued in the momentum $k$),
the inverse propagator has the form
\begin{align}
    \mathbf P^2 + m_\pi^2 \, \mathbf 1 + \mathbf\Sigma 
    \sim
    \mathbf{Z}^{-1/2}
    \left[
	(\mathbf P_\mu + \delta p_\mu)^2
	+ m_\pi^2 \, \mathbf 1
	+ \delta m_\pi^2
    \right]
    \mathbf{Z}^{-1/2}
    \,,
\label{propform}
\end{align}
where $m_\pi$ denotes the infinite volume pNGB mass,
$\delta m_\pi^2 \equiv \| (\delta m_\pi^2)^A_{\;\;B}\|$ is the FV mass correction,
and $\mathbf{Z} \equiv \|Z^A_{\;\; B}\|$ is a FV wavefunction renormalization factor.
The FV momentum shift $\delta p_\mu \equiv \|(\delta p_\mu)^A_{\;\; B}\|$
is independent of the external
momentum $k$.
This correction leads to a change in the relation between
4-momentum and 4-velocity, and hence is naturally
regarded as causing a \emph{velocity shift}.%
\footnote
    {%
    This shift has previously been
    described as a renormalization of the twist angle
    \cite{Colangelo:2016wgs},
    and as a renormalization of the field momentum in
    \cite{Jiang:2006gna};
    see also \cite{Bijnens:2014yya} for another perspective.
    However, the twist angles determine the quantization of momenta in the box,
    and this quantization cannot be affected by interactions.
    Thus, we find more helpful our above characterization
    as a ``velocity shift''.
    The FV correction leading to the velocity shift arises in
    a calculation of the influence of interactions on the energy
    of an excitation, not its momentum.
    }
Neglecting (for simplicity) the FV mass shift,
the group 4-velocity $u^\mu$ of a pion is given by
\begin{equation}
    u = \frac {P + \delta p}{m_{\pi}} = u_\infty + \delta u \,,
\end{equation}
and deviates from the infinite volume result
$u_\infty \equiv P/m_\pi$.
The velocity shift $\delta u = \delta p/m_\pi$
vanishes with periodic or anti-periodic quark boundary conditions,
as well as in the special case of $\mathbb{Z}_{\Nf}$-symmetric
twisted boundary conditions (discussed in the next section),
but is otherwise non-zero.

The sums appearing in the FV self-energy \eqref{1loopFVdiag},
suitably regularized with their infinite volume limits subtracted,
can be computed using the regularized tadpole
$T(s,\alpha_\mu)$ given in Eq.~\eqref{eq:tadpolesum}
plus one related sum,
\begin{equation}
    T_\mu(s,\alpha_\mu)
    \equiv
    \sum_{n_\mu} \> \frac{N_\mu}{\left( N \cdot N + m_\pi^2\right)^s}
    =
    \frac{L_\mu}{2(s{-}1)} \, \frac{\partial}{\partial \alpha_\mu} \,
    T(s{-}1, \alpha_\mu)\,.
\label{linearsum}
\end{equation}
With these results, the mass shift, wavefunction renormalization,
and velocity shift corrections to the pion propagator can readily be
expressed in terms of the holonomy.
One finds that
the velocity shift correction is given by
\begin{align}
  (\delta u_\mu)^{A}{}_{B}
    &=
    \frac{m_\pi}{2(4\pi f_\pi)^2} \!\sum_{C,D}
    \left[ \tfrac{2}{\Nf}
	\big(
        \delta^A_{\;\;C} \, \delta^D_{\;\;B}
	    - g^{AD} \, g_{BC} \big) 
            + d^A_{\;\;CE} \, d_B^{\;\;D E}
	    - d^E_{\;\;BC} \, d_E^{\;\;AD}
    \right] \!\!
    \nonumber
    \\ &\qquad\qquad {} \times
    \sum_{n_\mu\in \mathbb{Z}^4}\!\!\!{\vphantom{\Big|}}^\prime \,
    \frac{n_\mu L_\mu}{|nL|^2}K_2(m_\pi|nL|) \>
    i \>
    [\Omega^n]^C_{\;\; D} ,
\label{twistrenrom}
\end{align}
where repeated Lorentz indices are not summed in $n_\mu L_\mu$.
The symmetries of $d$-symbols,
the vanishing twist angles of neutral pNGBs,
and the fact that 
the adjoint representation holonomy is diagonal in our basis,
together imply that the velocity shift vanishes for neutral pNGBs
(when $A = \bar A$ and $B = \bar B$).
Generic non-zero twist angles do lead to velocity shifts for
charged pNGBs (for which $A \ne \bar A$), but
the unbroken $U(1)^{\Nf-1}_V$ Cartan flavor symmetry implies that
the velocity shift can only be non-zero when $A$ and $B$ are
charge conjugates of each other.
Hence
$
    (\delta u_\mu)^A_{\;\;B} = (\delta u_\mu)^A_{\;\;A} \> \delta^A_{\;\; B}
$,
and no mixing among pNGBs is induced by the velocity shift.
For any choice of twisted boundary conditions,
the flavor-averaged velocity shift vanishes identically,
\begin{equation}
    \overline {\delta u_\mu}
    \equiv
    \frac {\trA [\delta u_\mu]}{\Nf^2{-}1} 
    =
    \frac 1{\Nf^2{-}1} \sum_A \> (\delta u_\mu)^A_{\;\;A}
    = 0 \,.
\end{equation}

The mass corrections to individual pNGBs can similarly be expressed
in terms of the holonomy,
\begin{align}
    (\delta m_\pi^2)^A_{\;\;B}
    &=
    \frac{m_\pi^4}{2(4\pi f_\pi)^2} \!\sum_{C,D}
    \Bigl[
	\tfrac{2}{\Nf}
        \big(\delta^A_{\;\;B} \, \delta^D_{\;\;C}
        - \delta^A_{\;\;C} \, \delta^D_{\;\;B}
        - g^{AD} \, g_{BC}\big)
    \nonumber
    \\ &\qquad\qquad
        {}
	+ d^A_{\;\;BE}\, d_C^{\;\;DE}
	- d^A_{\;\;CE}\, d_B^{\;\;DE}
	- d^E_{\;\;BC}\, d_E^{\;\;AD}
    \Bigr]
    \sum_{n_\mu\in \mathbb{Z}^4}\!\!\!{\vphantom{\Big|}}^\prime \>
    \frac{K_1( m_\pi |nL|)}{m_\pi |nL|} \>
    [\Omega^n]^C_{\;\;D} \,.
    \label{nonsingletmassshift}
\end{align}
Although not obvious from Eq.~\eqref{nonsingletmassshift}, all dependence on the twist angles cancels in the mass shift for charged pNGBs.  We have verified this via explicit evaluation for $\Nf \le 10$, but do not have a general argument.  The masses of all the charged pNGBs remain degenerate, for any choice of the holonomies.
The mass shift $\delta m_\pi^2$ can be non-diagonal only within the subspace
of neutral pNGBs.
In other words, flavor twisted boundary conditions generically induce mixing
among neutral pNGBS, but cannot mix pNGBs which are
charged under $U(1)^{\Nf-1}_V$.  The mixing vanishes with periodic and anti-periodic boundary conditions, as well as with $\mathbb{Z}_{\Nf}$-symmetric boundary conditions. 

The flavor averaged pNGB mass shift simplifies dramatically
and has the expected dependence on adjoint traces of the holonomy,
\begin{align}
    \overline{\delta m_\pi^2}
    \equiv
    \frac{\trA\!\big[ \delta m_\pi^2\big]}{\Nf^2{-}1}
    &=
    \frac{(m_\pi^2/4\pi f_\pi)^2}{\Nf\,(\Nf^2{-}1)}
    \sum_{n_\mu\in \mathbb{Z}^4}\!\!\!{\vphantom{\Big|}}^\prime \>
    \frac{K_1\big( m_\pi |nL| \big)}{m_\pi |nL|} \>
    \trA[\Omega^n] \,.
\label{mass}
\end{align}
All the conclusions regarding vanishing-adjoint BCs
detailed above for the free energy also hold for the flavor averaged pNGB mass.
In particular, vanishing-adjoint BCs applied only in the
time direction will exponentially reduce
the one-loop finite $\beta$ artifacts to
$O(e^{-(\Nf+1)m_\pi\beta})$.
For very light pions,
when $m_\pi/f \ll 1$ and $m_\pi\beta \ll 1$,
vanishing-adjoint boundary conditions in time
lead to a finite-$\beta$ mass shift,
\begin{equation}
    \overline{\delta m_\pi^2}
    \sim
    \frac {m_\pi^2/(4\pi f_\pi)^2}{\Nf \, (\Nf{+}1)^2 \beta^2} \,,
\label{eq:massshift}
\end{equation}
which is suppressed by $1/(\Nf{+}1)^2$
[in addition to the $1/(\Nf f_\pi^2)$ factor which scales
as $1/(\Nc\Nf)$ for large $\Nc$ and $\Nf$].
More generally,
applying vanishing-adjoint BCs in all spacetime directions removes
the leading $n\cdot n = 1$ shell of finite volume corrections, thereby
reducing FV artifacts to $O(e^{-\sqrt{2} m_\pi L})$ (for a symmetric box)
in all directions.

Specializing the above results to the specific case of $\Nf = 3$,
one finds
\begin{subequations}\label{eq:vshift3}
\begin{align}
  (\delta u_\mu)_{\pi^{\pm}} &=
  \pm\frac{m_{\pi}}{32\pi^2f_{\pi}^2}
  \sum_{n_\mu\in \mathbb{Z}^4}\!\!\!{\vphantom{\Big|}}^\prime
      \left[
        \sin (n \cdot \alpha_{K^0})
        -\sin (n \cdot \alpha_{K^{+}})
        -2 \sin (n \cdot \alpha_{\pi^{+}})
      \right]
      \frac{n_\mu L_\mu}{|nL|^2}K_2(m_\pi|nL|) ,
\\
      (\delta u_\mu)_{K^{\pm}} &=
      \mp\frac{m_{\pi}}{32\pi^2f_{\pi}^2}
      \sum_{n_\mu\in \mathbb{Z}^4}\!\!\!{\vphantom{\Big|}}^\prime
      \left[
	  \sin (n \cdot \alpha_{K^0})
	  +2\sin (n \cdot \alpha_{K^{+}})
	  + \sin (n \cdot \alpha_{\pi^{+}})
      \right]\,
      \frac{n_\mu L_\mu}{|nL|^2}K_2(m_\pi|nL|),
\\
      (\delta u_\mu)_{K^{0}, \bar{K}^{0}} &=
      \mp\frac{m_{\pi}}{32\pi^2f_{\pi}^2}
      \sum_{n_\mu\in \mathbb{Z}^4}\!\!\!{\vphantom{\Big|}}^\prime
      \left[
	  2 \sin (n \cdot \alpha_{K^0})
	  + \sin (n \cdot \alpha_{K^{+}})
	  - \sin (n \cdot \alpha_{\pi^{+}})
      \right]\,
      \frac{n_\mu L_\mu}{|nL|^2}K_2(m_\pi|nL|),
\end{align} 
\end{subequations}
where again repeated spacetime indices are not summed in $n_\mu L_\mu$.
For $\Nf=3$,
the charged meson mass shift matrix turns out not depend on the twist angles
whatsoever,
and thus is the same for all the charged mesons,
\begin{align}
    \delta m_{\pi}^2\big|_{\textrm{charged pNGBs}}
    =
    \frac{m_{\pi}^4}{48 \pi^2 f_{\pi}^2}
    \sum_{n_{\mu} \in \mathbb{Z}^4}\!\!\!{\vphantom{\Big|}}^\prime \>
    \frac{K_1( m_\pi |nL|)}{m_\pi |nL|} \,,
\label{eq:chargedNGBmassShift}
\end{align}
while the mass shift in the neutral ${\pi^0,\eta}$ pNGB sector
takes the form
\begin{align}
    \delta m_{\pi}^2\big|_{\pi^0,\eta}
    &=
    \frac{m_{\pi}^4}{48 \pi^2 f_{\pi}^2} \!
    \sum_{n_{\mu} \in \mathbb{Z}^4}\!\!\!{\vphantom{\Big|}}^\prime  \>
    \left(\begin{array}{cc}
      \scriptstyle
	  {-2+ 3 \cos ( n \cdot \alpha_{\pi^{+}})}
      &
      \scriptstyle
	  {\sqrt{3}\, 
	  \left[
	      \cos ( n \cdot \alpha_{K^{+}})
	      -\cos ( n \cdot \alpha_{K^{0}})
	  \right]} 
      \\
      \scriptstyle
	  {\sqrt{3}\,
	  \left[
	      \cos ( n \cdot \alpha_{K^{+}})
	      -\cos ( n \cdot \alpha_{K^{0}})
	  \right]}
      & 
      \scriptstyle
	  {\;-2+ 2\cos (n\cdot\alpha_{K^0} )
	      + 2\cos (n\cdot\alpha_{K^{+}})
	      -\cos (n\cdot\alpha_{\pi^{+}}) }
    \end{array}\right)
\nonumber
\\ &\hspace*{1in}{}\times
    \frac{K_1( m_\pi |nL|)}{m_\pi |nL|} \,.
\label{eq:PiEtaMixing}
\end{align}
Here $\alpha^{\mu}_{\pi^{+}} = \alpha^{\mu}_{1}-\alpha^{\mu}_{2}$,
$\alpha^{\mu}_{K^{+}} = \alpha^{\mu}_{1}-\alpha^{\mu}_{3}$,
$\alpha^{\mu}_{K^{0}} = \alpha^{\mu}_{2}-\alpha^{\mu}_{3}$
are the twists for the $\pi^{+}$, $K^{+}$, and $K^{0}$
particles written in terms of the fundamental representation quark twists.
The neutral mass shift matrix \eqref{eq:PiEtaMixing}
reduces to a multiple of the identity for vanishing twist angles,
$\alpha^{\mu}_{\pi^{+}} = \alpha^{\mu}_{K^{+}}=\alpha^{\mu}_{K^{0}}=0$, 
(corresponding to periodic (or antiperiodic) quark boundary conditions),
or when
$\alpha^{\mu}_{\pi^{+}} = -2\pi/3$,
$\alpha^{\mu}_{K^{+}}=-4\pi/3$, and
$\alpha^{\mu}_{K^{0}}= - 2\pi/3$, 
corresponding to $\mathbb{Z}_{3}$-symmetric boundary conditions on quarks.  
The off-diagonal mass shift terms, leading to $\pi^0$-$\eta$ mixing,
are non-zero for our vanishing-adjoint boundary conditions,
$\alpha^{\mu}_{\pi^{+}} = -\pi/2$,
$\alpha^{\mu}_{K^{+}}=\pi/2$,
$\alpha^{\mu}_{K^{0}}= \pi$.

Our explicit $\Nf=3$ expressions for the velocity shifts \eqref{eq:vshift3},
the charged meson mass shift \eqref{eq:chargedNGBmassShift},
and the diagonal elements of the neutral meson mass shift matrix
\eqref{eq:PiEtaMixing}
agree with the corresponding results in ref.~\cite{Colangelo:2016wgs}.
However, the authors of ref.~\cite{Colangelo:2016wgs}
did not discuss the off-diagonal terms associated with $\pi^0$-$\eta$ mixing.

\section{Higher loops and the $1/\Nf$ expansion}\label{sec:loops}

The basic observation underlying the utility of 
vanishing-adjoint-trace boundary conditions for suppression of
finite volume artifacts is the association 
between factors of $e^{-m_\pi L}$ and traces of the flavor holonomy
in the leading finite volume corrections to flavor singlet observables
(for sufficiently light pions in large boxes).
The leading FV corrections are associated with a single
excitation wrapping once around a compactified direction, and the
amplitude for such propagation will necessarily involve an exponential of
the particle mass times propagation distance, multiplied by
the holonomy appropriate for the flavor representation of the
particle.
As long as the lightest hadronic state transforms in the adjoint flavor
representation, then adjoint traces of the flavor holonomy will control
the leading finite volume effects in exactly the same manner as in our
explicit examples.
Reduction of $O(e^{-m_\pi L})$ finite volume 
effects to $O(e^{-\sqrt{2} m_\pi L})$ for flavor-singlet observables
in hypercubic volumes is a robust result
that holds at all orders in chiral EFT
(provided no other hadrons are lighter than $\sqrt{2}m_\pi$
and thus dominate the residual finite volume effects).

As we have seen, with a single relevant compactified direction
(e.g., $\beta \ll L_i$)
vanishing-adjoint twisted boundary conditions can
eliminate much more than just the leading winding number $\pm1$ terms;
they can remove FV artifacts involving arbitrary windings which are
non-zero modulo $\Nf{+}1$.
This was seen explicitly in our one-pion-loop results for the free energy
(\ref{eq:twistedfreeenergy}) and flavor-averaged pion mass
(\ref{mass})-\eqref{eq:massshift},
as well as the tree-level position space pion propagator
(\ref{eq:thermalnoise})-\eqref{eq:Cpichiral}.
However, this dramatic reduction in sub-leading finite volume artifacts,
eliminating all contributions from winding numbers $|n| = 2, {\cdots}, \Nf$,
is not a general result.
At higher loop orders there exist sub-leading finite volume
effects in which the equality between the total number of powers
of $e^{-m_\pi L}$ and the magnitude of the \emph{net} winding number
no longer holds.
This happens in contributions involving an excitation which
loops around the circle in one direction, interacts, and then
loops around the circle in the opposite direction.
However, these surviving sub-leading finite volume corrections,
involving winding numbers $1 < |n| \le \Nf$,
are suppressed by factors of $O(1/\Nf^2)$.
This suppression is best understood as an aspect of large $N$ volume
independence, which we briefly recap.

A general feature of $SU(\Nc)$ gauge theories, when compactified on a torus, 
is large $\Nc$ volume independence
\cite{Eguchi:1982nm,Bhanot:1982sh,Unsal:2010qh}.%
\footnote
    {%
    There is also a notion of large $N$ volume independence in
    vector-type field theories
    \cite{Dunne:2012ae,Dunne:2012zk,Dunne:2015ywa,Sulejmanpasic:2016llc}.
    }
One aspect of this phenomena,
for theories compactified on a circle of size $L$,
is that planar diagrams contributing to local single-trace
observables depend on the compactification size $L$ only via the
combination $\Nc L$,
provided the $\mathbb Z_{\Nc}$ center symmetry is unbroken
\cite{Eguchi:1982nm,
    GonzalezArroyo:1982hz,
    Gross:1982at,
    GonzalezArroyo:2010ss,
    Cherman:2014ofa}.
More generally, at a non-perturbative level,
the leading large $\Nc$ behavior of single trace expectation values,
or connected correlation functions of such operators,
is volume independent, with corrections to this limit vanishing
as $O(1/\Nc^2)$.%
\footnote
    {%
    See refs.~\cite{Unsal:2010qh} for a more precise characterization of the
    set of operators to which this statement applies.
    }

In our context, we have an $SU(\Nc)$ gauge theory with an $SU(\Nf)$
global flavor symmetry.
When focusing attention on flavor singlet observables,
one may equally well regard the theory as the limit of an
$SU(\Nc) \times SU(\Nf)$ gauge theory, in which the coupling
of the $SU(\Nf)$ gauge group is sent to zero.
The same structure of large $N$ volume independence applies to
such product gauge theories (when $\Nc$ and $\Nf$ both become large),
provided the $\mathbb Z_{\Nc} \times \mathbb Z_{\Nf}$ center symmetry is
unbroken.%
\footnote
    {%
    We assume that $\Nc$ is scaled with $\Nf$ in such a manner that
    the theory remains in an asymptotically free confining phase.
    }
In a perturbative context, this is the same as requiring
that the flavor holonomy around the compactified direction
have a $\mathbb Z_{\Nf}$ symmetric distribution of eigenvalues
(and likewise for the color holonomy).
In the limit of zero flavor coupling, this is the same as simply
choosing a non-dynamical $\mathbb Z_{\Nf}$-symmetric flavor holonomy:
\begin{align}
    \Omega_{\rm center-sym} =
    (-1)^{\Nf-1} \>
    \mathrm{diag}(1, \omega, \omega^2, \cdots, \omega^{\Nf-1})\,,
\label{eq:centersym}
\end{align} 
where $\omega \equiv e^{2\pi i/\Nf}$.
This is similar, but not identical, to our vanishing adjoint trace
holonomy (\ref{eq:vanishingadjoint}).
The $\mathbb Z_{\Nf}$-symmetric holonomy \eqref{eq:centersym}
sets to zero fundamental representation traces
with winding numbers of the holonomy which are non-zero modulo $\Nf$.
Adjoint representation traces do not vanish but are $1/\Nf^2$ suppressed relative to their typical $O(\Nf^2)$ scale,
\begin{equation}
    \trA \Omega_{\rm center-sym}^k = -1  \quad
    \mbox{if $k \bmod \Nf \ne 0$.}
\end{equation}
In other words, for large $\Nf$ and fixed winding numbers,
traces of the center-symmetric holonomy (\ref{eq:centersym})
and our vanishing-adjoint holonomy (\ref{eq:vanishingadjoint})
differ only by relative $O(1/\Nf^2)$ corrections.
Consequently, all the implications of large $N$ volume independence
equally apply to our vanishing adjoint boundary conditions, up to
relative $1/\Nf^2$ corrections.
In particular,
if one uses the center-symmetric twist (\ref{eq:centersym}) 
instead of our vanishing-adjoint twist (\ref{eq:vanishingadjoint}),
then the leading $O(e^{-m_\pi L})$ finite-volume corrections
will be reduced by a factor of $1/(\Nf^2{-}1)$, but not eliminated.
Indeed, for $\Nf=2$ this reduction of the finite-volume
artifacts by a factor of $3$ with center-symmetric boundary conditions was
observed in ref.~\cite{Briceno:2013hya}
(which referred to the center-symmetric
twist as the ``i-periodic" boundary condition).

The above discussion applies to an $S^1$ compactification.
As seen in earlier sections, if the theory is compactified on
a multidimensional torus with the same vanishing-adjoint twisted
boundary conditions in all directions, $\Omega_\mu = \Gamma$,
then one can eliminate
the leading exponential $O(e^{-m_{\pi}L})$
artifacts in flavor singlet observables, but not the next $O(e^{-\sqrt{2} m_{\pi}L})$ artifacts.
However, if $\Nf$ is a suitably chosen composite integer, then
it is possible to choose different (but commuting) twists for
each direction in a manner which, for large $\Nf$, independently
eliminates FV contributions involving windings up to a given order
about each direction, up to $O(1/\Nf^2)$ residuals.
By using a sequence of increasing composite values of $\Nf$,
with suitably chosen $\Nf$-dependent holonomies in each direction,
flavor singlet observables should exhibit volume independence in the
large $\Nf$ limit (with the ratio $\Nf/\Nc$ held fixed).
For example, if $\Nf = K^D$ for some integer $K > 1$, then
the $(\mathbb Z_K)^D$ symmetric choice
\begin{subequations}
\begin{align}
    \Omega_1 &= Z_K \otimes 1_K \otimes 1_K \otimes \cdots \otimes 1_K \,,
\\
    \Omega_2 &= 1_K \otimes Z_K \otimes 1_K \otimes \cdots \otimes 1_K \,,
\\
    &\vdots
\nonumber
\\
    \Omega_D &= \underbrace{1_K \otimes 1_K \otimes 1_K \otimes \cdots \otimes Z_K}_{D\;\rm factors} \,,%
\end{align}
\end{subequations}
with $1_K$ denoting a $K$-dimensional identity matrix and
$Z_K \equiv (-1)^{K-1} \mathrm{diag}(1,\gamma,\gamma^2,{\cdots},\gamma^{K-1})$
a diagonal matrix with all $K$'th roots of unity,
produces vanishing fundamental representation traces
unless \emph{all} winding number components are
multiples of $K$,
$
    \trF [\Omega^n] = 0
$
if any
$n_\mu \ne 0 \pmod K$.
Therefore, the corresponding adjoint representation traces are all
$\O(1)$,
$
    \trA [\Omega^n] = -1
$
if any $n_\mu \ne 0 \pmod K$,
and not $O(\Nf^2)$.
So, for example, if $\Nf = 2^4$, then for theories on $T^4$
the imposition of such flavor twisted boundary conditions will suppress
by a factor of 15 finite volume effects from all contributions involving
odd winding numbers about any dimension.
Although not helpful for QCD with three light quarks,
this idea should be useful in lattice studies of the
conformal window which explore the behavior of
gauge theories with large values of $\Nf$,
many of which are composite numbers.


\section {Discussion}  \label{sec:discussion}

In compactified theories, most
hadronic properties differ from their infinite volume values
by corrections which, in theories with a mass gap,
are exponentially dependent on the
mass of the lightest particle.
Appropriately chosen flavor-twisted boundary conditions
can remove these leading exponential corrections from flavor-singlet
observables, provided the lightest particle is a flavor non-singlet.
The possibility of achieving exponential reduction of finite volume effects 
with TBCs has been previously demonstrated for 
one- and two-baryon systems in ref.~\cite{Briceno:2013hya}.
This work describes how the vanishing-adjoint symmetry condition guarantees
exponential reduction of finite volume effects in
generic flavor-singlet observables.
We have explicitly demonstrated at one-loop that vanishing-adjoint boundary conditions
reduce finite temperature corrections to the free energy and the
flavor-averaged pNGB mass from $O(e^{-m_\pi\beta})$ to
$O(e^{-(\Nf+1)m_\pi\beta})$
when TBCs are imposed on just the time direction,
or more generally reduce finite volume corrections
from $O(e^{-m_\pi L})$ to $O(e^{-\sqrt{2}m_\pi L})$ if
vanishing-adjoint BCs are adopted in all directions in hypercubic volume.
Analogous results for FV artifact reduction of $\Nf=2$ flavor-averaged
baryon masses with vanishing-adjoint BCs were previously observed in
ref.~\cite{Briceno:2013hya}.

The utility of vanishing-adjoint boundary conditions to reduce
finite volume artifacts is by no means limited to the observables we have explicitly calculated,
or to theories near the chiral limit.
What is required is that one considers a theory where the lightest
hadronic state transforms in the adjoint representation, with the theory
compactified in a box which is large compared to the Compton wavelength
of this excitation.
Our results may be understood as arising from background
flavor gauge invariance, which ensures that flavor singlet observables
can only depend on traces of the flavor holonomy.
Since pNGBs transform as flavor adjoints, all $O(e^{-m_\pi L})$
FV corrections arising from pNGB loops will depend on the adjoint trace
of the holonomy, and are removed by vanishing-adjoint boundary conditions.
For hypercubic compactifications, with identical TBCs in all directions,
finite volume artifacts are reduced to $O(e^{-\sqrt{2}m_\pi L})$
(assuming no other particles have masses below $\sqrt 2 \, m_\pi$).
This same phenomena should apply to generic flavor-singlet observables,
such as other flavor-averaged masses and matrix elements, as long as
the observable of interest is not probing multi-particle states very
near or above scattering thresholds.  

For scattering or near-threshold bound states,
a small relative momentum or binding momentum,
instead of the pNGB mass, may control the leading FV effects
\cite{Luscher:1985dn,Beane:2003da,Luscher:1986pf}.
The deuteron binding energy provides an explicit example
where examination of the flavor twist dependence \cite{Briceno:2013hya}
has shown that vanishing-adjoint BCs do not remove the leading FV
artifacts which depend exponentially on the deuteron binding momentum.
Other choices of TBCs can be used to reduce FV artifacts
specifically in deuteron binding energy
calculations~\cite{Briceno:2013hya},
and extensions of these TBCs for other systems are being
explored~\cite{Korber:2015rce}.

It is important to note that the effect of TBCs on binding energies
and scattering parameters extracted from lattice simulations
is calculable and has already been explored for many systems.
The use of vanishing-adjoint BCs does not preclude scattering
parameter extraction as long as TBCs are properly included in all
FV quantization conditions.
Removal of $O(e^{-m_\pi L_\mu})$ artifacts may be of significant
utility for extractions of scattering parameters that measure power
law volume dependence under the assumption that $O(e^{-m_\pi L_\mu})$
artifacts are negligible.
Detailed studies of vanishing-adjoint boundary conditions in the two-body sector
will be needed to understand these effects and are left to future
work.

\acknowledgements
We are grateful to  S.~R.~Beane, D.~B.~Kaplan, M.~J.~Savage, E.~Shaghoulian, S.~R.~Sharpe,
B.~C.~Tiburzi, and E.~Witten for helpful discussions.
This work was supported in part by the U.~S.~Department of Energy
under grants
DE-FG02-00ER-41132 (A.C.),
DE-FG02-04ER41338 (S.S.),
DE-FG02-00ER41132 (M.L.W) and
DE-SC0011637 (L.G.Y).

\bibliography{small_circle} 

\end{document}